\documentstyle[referee,epsfig]{mn}

\newcommand{\be}{\begin{equation}}
\newcommand{\ee}{\end{equation}}

\begin{document}

\title[Are there cosmological evolution trends on Gamma-Ray Burst features?]
{Are there cosmological evolution trends on Gamma-Ray Burst
features?}

\author[D.M. Wei,  W.H. Gao]
{D.M. Wei$^{1,2}$,  W.H. Gao$^{1,2}$ \\
$^{1}$ Purple Mountain Observatory, Chinese Academy of Sciences, Nanjing, 210008, China\\
$^{2}$ National Astronomical Observatories, Chinese Academy of Sciences, China\\
}

\maketitle

\noindent{\bf submitted to MNRAS Main Journal} \vspace{5mm}

\noindent{All correspondence please send to:} \vspace{5mm}

\noindent{D.M. Wei}\\
Purple Mountain Observatory\\
Chinese Academy of Sciences\\
Nanjing, 210008\\
P.R. China

\vspace{5mm}
\noindent{email: dmwei@pmo.ac.cn}\\
fax: 8625-3307381

\newpage

\begin{abstract}

The variability of gamma-ray burst (GRB) is thought to be correlated
with its absolute peak luminosity, and this relation had been used to
derive an estimate of the redshifts of GRBs. Recently Amati et al.
presented the results of spectral and energetic properties of several
GRBs with known redshifts. Here we analyse the properties of two group
GRBs, one group with known redshift from afterglow observation, and
another group with redshift derived from the luminosity - variability
relation. We study the redshift dependence of various GRBs features in
their cosmological rest frames, including the burst duration, the
isotropic luminosity and radiated energy, and the peak energy $E_p$ of
$\nu F_\nu$ spectra. We find that, for these two group GRBs, their
properties are all redshift dependent, i.e. their intrinsic duration,
luminosity, radiated energy and peak energy $E_p$, are all correlated
with the redshift, which means that there are cosmological evolution
effects on gamma-ray bursts features, and this can provide an
interesting clue to the nature of GRBs. If this is true, then the
results also imply that the redshift derived from the luminosity -
variability relation may be reliable.

\end{abstract}


\newpage

\section{Introduction}

The study of gamma-ray bursts (GRBs) afterglows has enable the
measurement of their distances, so far GRBs are known as an explosive
phenomenon occurring at cosmological distances, emitting large amount
of energy mostly in the gamma-ray range (see, e.g.  Piran 1999; Cheng
\& Lu 2001 for a review). So GRBs can provide useful information about
the early epochs in the history of the universe.

Although a great achievements have been made about the GRB afterglows,
we still know little about the origin of gamma-ray bursts, the reason
is that the GRBs with known redshifts are relatively rare, now there
are only about 20 GRBs with known redshifts, so it is difficult to do
some statistics about GRBs features, such as their luminosity function,
duration distribution, etc.. However, two important correlations have
been discovered, i.e. between the degree of variability of the
gamma-ray burst light curve and the GRB luminosity (Ramirez-Ruiz \&
Fenimore 1999; Feminore \& Ramirez-Ruiz 2001), and between the
differential time lags for the arrival of burst pulses at different
energies and the GRB luminosity (Norris, Marani \& Bonnell 2000),
although these correlations are still tentative, they offer the
possibility to derive independent estimates of the redshifts of GRBs.

Recently Amati et al. (2002) have reported the spectral and
energetic properties of several GRBs with known redshifts, these
bursts were all detected by BeppoSAX and have good-quality
time-integrated spectra. In addition, Lloyd-Ronning \&
Ramirez-Ruiz (2002) have found that bursts with highly variable
light curves have greater $\nu F_\nu$ spectral peak energies in
their cosmological rest frames. These results reinforce the
validity of the redshift estimates derived from the luminosity -
variability relation. Here we will discuss the properties of two
group GRBs, one group includes the bursts with known redshifts and
well-defined spectra detected by BeppoSAX, and another group
consists of bursts whose redshifts are derived from the luminosity
- variability relation. We will show that the properties of these
two group GRBs are all correlated with redshift, which suggests
that the luminosity - variability relation may be reliable, and
furthermore the GRBs' features are redshift dependent.

\section{The Properties of Two Group GRBs}

Amati et al. (2002) have analysed the spectral properties of the X-ray
and gamma-ray emission from GRBs with known redshifts. Those bursts
were all detected by BeppoSAX satellite and have good-quality spectra.
The extension of the spectral analysis to the X-ray energy band allows
a better determination of the continuum spectrum, reducing the bias in
the measurement of the spectral slope below the peak energy $E_p$ of
$\nu F_\nu$ spectra. In their sample total 12 gamma-ray bursts were
included, however among them, there are three bursts (GRB980326,
GRB980329 and GRB000214) whose redshifts are not determined accurately,
only redshift intervals are given, so in our one group we ignore these
three bursts, and contains the other 9 bursts with firm redshifts.

GRB temporal profiles are so complicated that, at first sight, their
behavior obeys no simple rule. However, several authors have suggested
that there may be correlation between the properties of burst time
structure and burst luminosity (e.g. Feminore \& Ramirez-Ruiz 2001;
Reichart et al. 2001; Norris, Marani \& Bonnell 2000). In particular,
Fenimore \& Ramirez-Ruiz (2001) explored the possibility of using the
"spikiness" of the time structure, combined with the observed flux, to
obtain the GRB redshifts. They have analysed several hundred long and
bright bursts and derived the redshifts and luminosities for 220 BATSE
bursts. Lloyd-Ronning \& Ramirez-Ruiz (2002) used total 159 bursts from
the above 220 GRBs to investigate the dependence of the burst spectra
on variability. The observed spectra of these 159 bursts can be well
characterized by the Band function (Band et al. 1993), defined by a low
energy spectral index $\alpha$, a high energy spectral index $\beta$,
and a peak energy of the $\nu F_\nu$ spectra $E_p$. So our second group
consists of these 159 bursts with known peak energy, and the redshifts
of them are derived from the luminosity - variability relation
(Feminore \& Ramirez-Ruiz 2001).

Fig.1 gives the GRB duration in their cosmological rest frame,
$T'=T/(1+z)$, versus the redshift, where $T'$ is the intrinsic duration
and $T$ is the observed duration of a burst. The filled circles are
bursts with secure redshifts estimates, while the empty triangles are
bursts in which the redshifts are derived using the luminosity -
variability distance indicator. Fig.2 and Fig.3 show the luminosity and
radiated energy of GRBs in their cosmological rest frame versus the
redshift, Fig.4 shows the relation between the peak energy $E_p$ of the
$\nu F_\nu$ spectra in their cosmological rest frame and the redshift.

From Fig.1 we see that although the distribution of the burst
duration is somewhat scatter, there is still a clear trend that
the intrinsic duration decreases with the redshift, if we fit a
power law to the data, we find $T'\propto (1+z)^{-0.85\pm 0.08}$.
Fig.2 and Fig.3 show that the isotropic luminosity and radiated
energy have a positive correlation with the redshift, excluding
the 4 bursts with smallest redshifts, we have $L \propto
(1+z)^{2.5\pm 0.1}$, $E\propto (1+z)^{1.62\pm 0.14}$. Fig.4 shows
that the peak energy of $\nu F_\nu$ spectra also has a positive
correlation with the redshift, the power law fit is $E_p \propto
(1+z)^{0.76\pm 0.07}$.

However, we know that the BATSE GRB sample is flux truncated, i.e. only
those bursts whose flux exceeding the threshold flux can be detected,
this effect can give rise to several apparent correlations. For
example, the burst duration (e.g. $T_{90}$) are defined by the time
interval over which $90\%$ of the fluence is detected. High-z bursts
are fainter, therefore some faint components of an intrinsic long burst
could have been cut out by the flux threshold effect, so that their
apparent durations are shortened. For another example, the correlation
trend in both Fog.2 and Fig.3 are also contaminated by the flux
truncation effect, as has been demonstrated by Lloyd-Ronning et
al.(2002). So we need some way to estimate the true correlations
between parameters that suffering from flux truncation.

Here we use the Monte Carlo simulation method. We simulate the
burst sample, assuming their luminosity satisfy $L =L_0
(1+z)^{\alpha}$, and their redshifts are distributed uniformly
between 0 and 10. In our simulation, we take the value $L_0
=3\times 10^{51}$ erg s$^{-1}$, the flux threshold is $10^{-7}$
erg cm$^{-2}$ s$^{-1}$, and the Hubble constant $H_0 =65$ km
s$^{-1}$ Mpc$^{-1}$, $\Omega_m=0.3$, $\Omega_\Lambda =0.7$. Under
this circumstance, we produce 300 bursts in original sample, and
find that 100 bursts of them with flux lower than the threshold,
so there are 200 bursts in our "observed" sample. Fig.5 shows our
simulation results, where the circles are the 300 bursts produced
by simulation, and the solid line is the flux threshold for BATSE.
We find that, when $\alpha=1.7\pm 0.5$, there is the relation $L
\propto (1+z)^{2.50\pm 0.08}$, which is consistent with the
observed value. So we see that the flux truncation has great
effect on the correlation coefficient.

In order to discuss this flux truncation effect on the burst
duration, one needs to know the intrinsic pulse profile. For
simplicity, here we assume the intrinsic pulse profile $L=L_{\rm
p} e^{-t^2/t_0^2}$, since only the component with $L\geq L_{\rm
lim}$ can be observed (where $L_{\rm lim}$ is the limit luminosity
corresponding to the flux threshold), so we can get the observed
pulse duration $T_{\star}=t_0(\ln \frac{L_{\rm p}}{L_{\rm
lim}})^{1/2}$. The intrinsic pulse duration should be proportional
to $t_0$, which is assumed to have the form $t_0 \propto
(1+z)^{\beta}$, $L_{\rm p}$ is assumed to take the form described
above, $L_{\rm p}\propto (1+z)^{1.7\pm 0.5}$. As above, we produce
a GRB sample with 300 bursts, 100 bursts of them are not
"observed" due to their low flux, so the remaining 200 bursts
consists of our "observed" sample. We find that when we take the
value of $\beta = 0.45 \pm 0.20$, the "observed" sample has the
relation $T_{\star}\propto (1+z)^{-0.84\pm 0.14}$, which is
compatible with the observed value.

Fig.6 shows the relation between the intrinsic peak energy of the
$\nu F_\nu$ spectra and the luminosity for group one, the solid
line represents the relation $E_p \propto L^{1/2}$. Fig.7 gives
the intrinsic peak energy of  $\nu F_\nu$ spectra versus the
luminosity for group two, the two solid lines represent the
relation $E_p \propto L^{1/2}$. We see that the relation $E_p
\propto L^{1/2}$ can account for the observed data quite well.

\begin{figure}
\centerline{\psfig{file=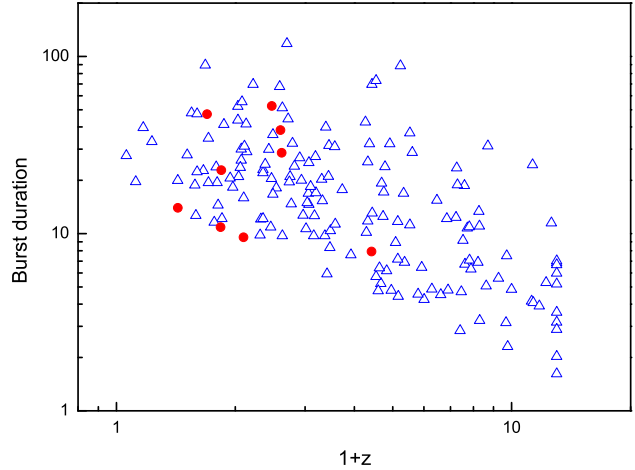,width=0.6\textwidth}} \caption{The
duration of GRBs in their cosmological rest frame versus the burst
redshift. The filled circles are bursts with secure redshifts
estimates, while the empty triangles are bursts in which the redshifts
are derived using the luminosity - variability distance indicator.}
\end{figure}

\begin{figure}
\centerline{\psfig{file=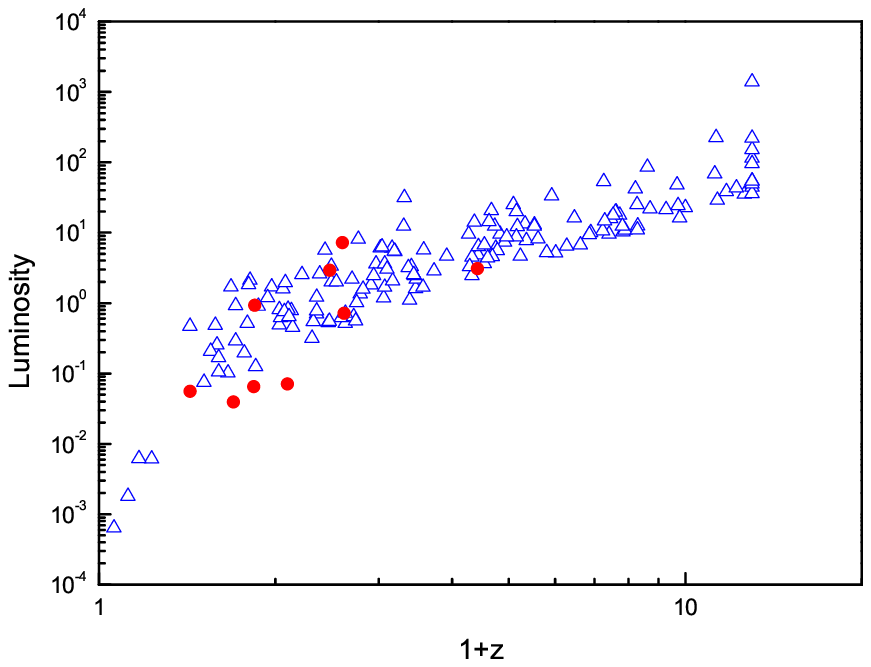,width=0.6\textwidth}} \caption{The
luminosity of GRBs in their cosmological rest frame versus the burst
redshift. The filled circles are bursts with secure redshifts
estimates, while the empty triangles are bursts in which the redshifts
are derived using the luminosity - variability relation.}
\end{figure}

\begin{figure}
\centerline{\psfig{file=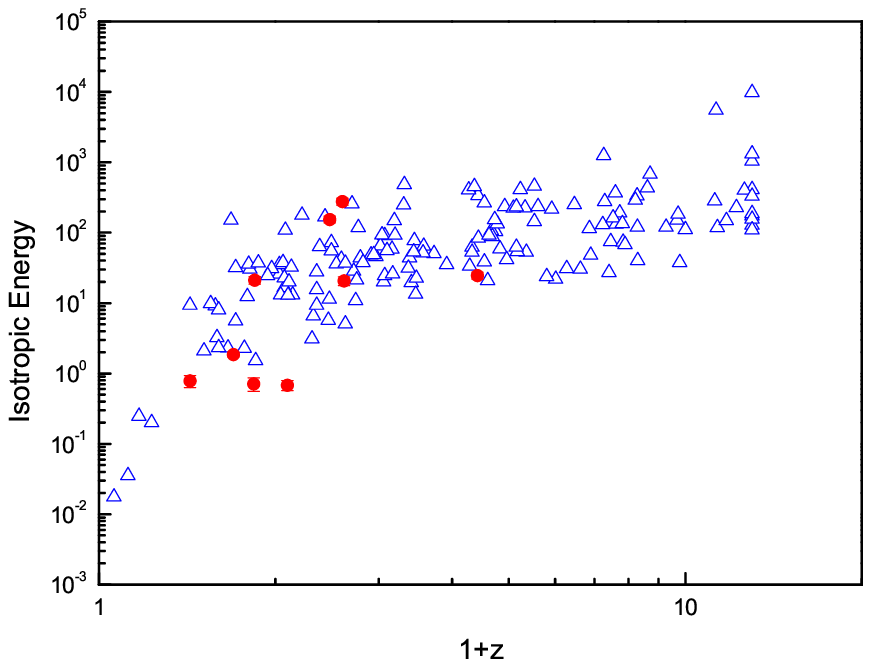,width=0.5\textwidth}} \caption{The
radiated energy of GRBs in their cosmological rest frame versus the
burst redshift. The filled circles are bursts with secure redshifts
estimates, while the empty triangles are bursts in which the redshifts
are derived using the luminosity - variability relation.}
\end{figure}

\begin{figure}
\centerline{\psfig{file=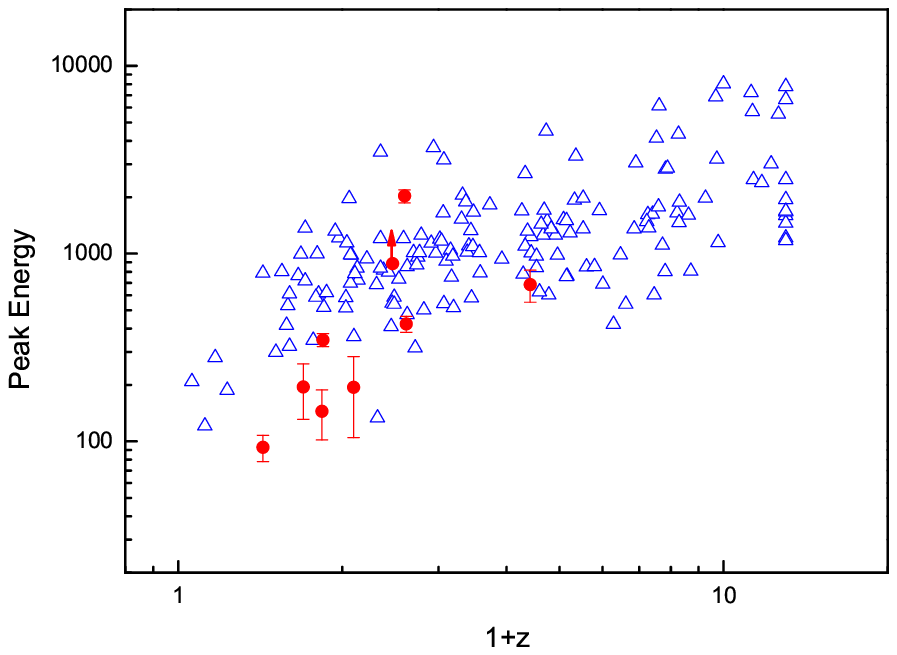,width=0.6\textwidth}} \caption{The
peak energy of $\nu F_\nu$ spectra of GRBs in their cosmological rest
frame versus the burst redshift. The filled circles are bursts with
secure redshifts estimates, while the empty triangles are bursts in
which the redshifts are derived using the luminosity - variability
relation.}
\end{figure}

\begin{figure}
\centerline{\psfig{file=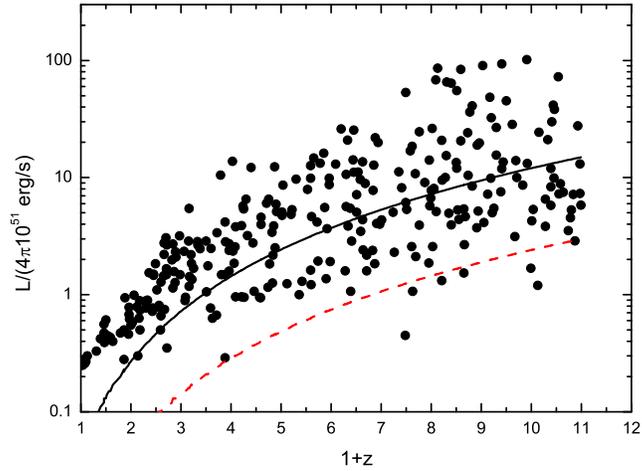,width=0.6\textwidth}}
\caption{Our simulation results, where the circles are the 300
bursts produced by the simulation, the solid line is the flux
threshold for BATSE, and the dashed line is the flux threshold for
Swift.}
\end{figure}

\begin{figure}
\centerline{\psfig{file=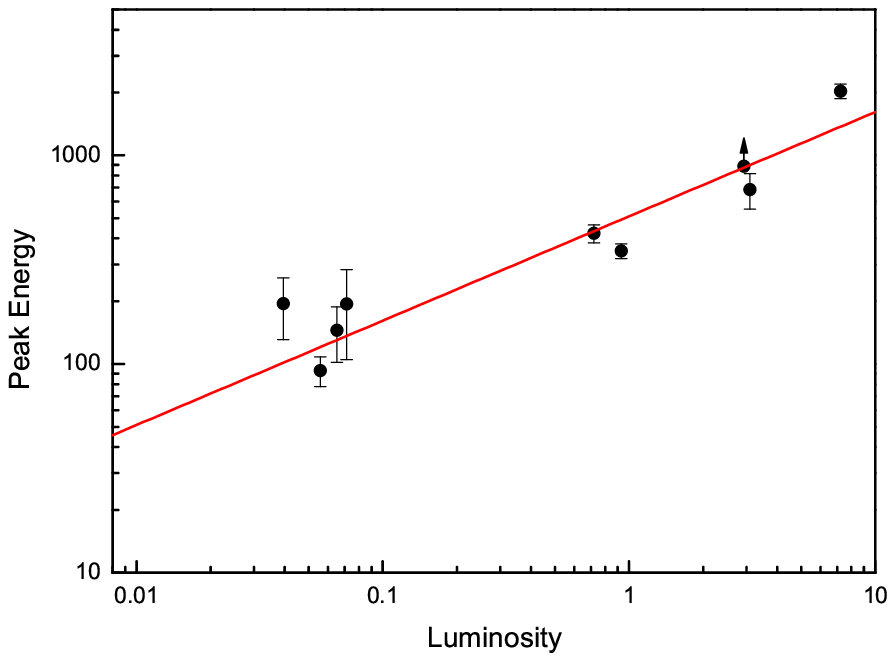,width=0.6\textwidth}}
\caption{The peak energy of $\nu F_\nu$ spectra of group one GRBs
in their cosmological rest frame versus the burst luminosity. The
solid line represents the relation $E_p \propto L^{1/2}$.}
\end{figure}

\begin{figure}
\centerline{\psfig{file=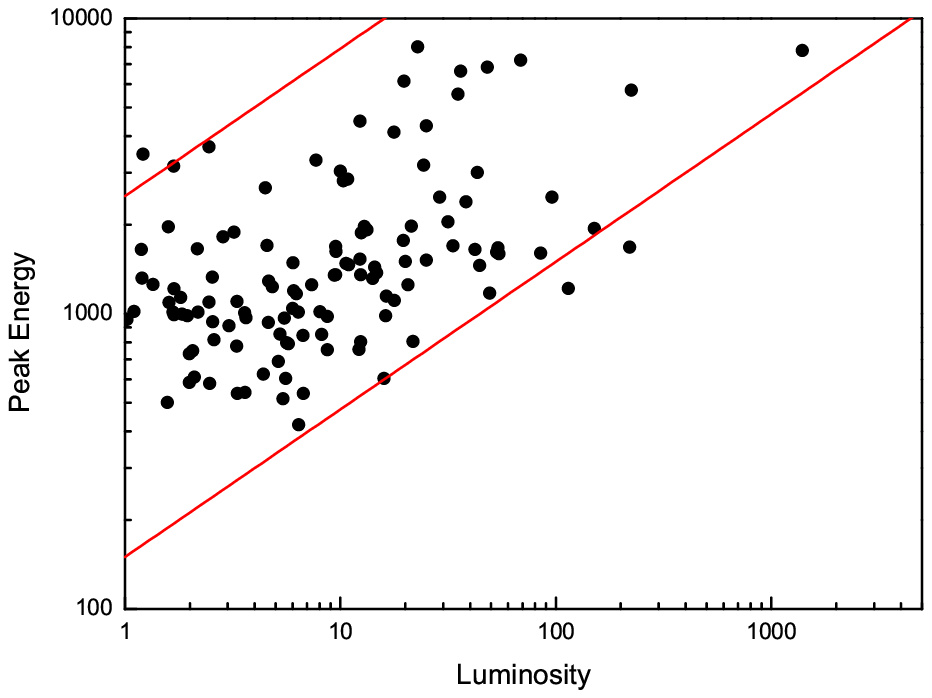,width=0.6\textwidth}}
\caption{The peak energy of $\nu F_\nu$ spectra of group two GRBs
in their cosmological rest frame versus the burst luminosity. The
two solid lines represent the relation $E_p \propto L^{1/2}$.}
\end{figure}

\section{Discussion and conclusion}

In previous section we have shown that the properties of our two
group GRBs, one group includes 9 GRBs with secure redshifts
derived from the afterglow observation,  another group consists of
159 GRBs with known peak energy and their redshifts are derived
from the luminosity - variability relation, are all correlated
with the redshifts, which reinforces the validity of the
luminosity - variability distance indicator. Our results further
support the conclusion obtained by Lloyd-Ronning \& Ramirez-Ruiz
(2002), they found that there is positive correlation between the
peak energy in the cosmological rest frame and the variability for
gamma-ray bursts whose redshifts are derived either from optical
spectral features or from the luminosity - variability distance
indicator.

Since the BATSE GRB sample is flux truncated, i.e. only those
bursts with flux exceeding the threshold flux can be detected, so
it is necessary to discuss this selection effects on the
statistical results. Here we use the Monte Carlo simulation
method. We find that, after correcting the selection effects, the
cosmological evolution trends are still exist, although they are
much shallower than the trends found in the face values of the
data. For example, if ignored the flux truncation effect, we
obtained the relation $L \propto (1+z)^{2.5\pm 0.1}$, while when
the selection effects are taken into account, we got the relation
$L \propto (1+z)^{1.7\pm 0.5}$, which is consistent with the value
$L \propto (1+z)^{1.4\pm 0.5}$ obtained by Lloyd-Ronning et al.
(2002).

Fig.5 illustrates our simulated results, where the circles are the
300 bursts produced by the simulation, the solid line is the flux
threshold for BATSE, and the dashed line is the flux threshold for
Swift. From this figure it is obvious that the property of the
"observed" sample should depend on the adopted flux threshold, for
different flux threshold their statistical properties are
different. For example, when taking the BATSE flux threshold,
there are 100 bursts with flux lower than the threshold, while
when taking the Swift flux threshold, there are only 7 bursts with
flux lower than the threshold, so in this case the observed $L-z$
relation should close to the true $L-z$ relation. Therefore we
expect that the luminosity - redshift relation for bursts observed
by Swift should be shallower than that for bursts observed by
BATSE.

Fig.6 and Fig.7 show that there is a good correlation between the
peak energy and luminosity for both group GRBs, and the relation
$E_p \propto L^{1/2}$ can account for the observed data quite
well. Up to now the location of the GRB emission site is still
unsettled, although the internal shock model is thought to be more
reasonable than the external shock model. Zhang \& Meszaros
(2002b) analysed various fireball models within a unified picture
and investigated the $E_p$ predictions of different models. It is
known that for internal shock model, if the GRB bulk Lorentz
factors are not correlated with the luminosities, then there is
the relation $E_p \propto L^{1/2}$, while for external shock model
$E_p \propto \Gamma ^4$, where $\Gamma$ is the shock Lorentz
factor. So our results suggest that the gamma-ray burst emission
are more likely from the internal shock.

Frail et al. have discussed the afterglow properties of several
GRBs with known redshifts, they assumed that the breaks in the
afterglow light curves are caused by the sideways expansion of the
jet, and then they concluded that the GRB emission energy is
nearly a constant, $E\sim 5\times 10^{50}$ ergs (Frail et al.
2001). However, Fig.3 shows that the isotropic radiated energy
increases with redshift, so if the conclusion of Frail et al. is
true, then we notice that the jet opening angle must decrease with
the redshift, as shown by Fig.8, where the data points are taken
from the paper of Frail et al. (2001). This point is very
interesting, since it can put constraints on the central engines
of GRBs. Of course, this phenomena can also be explained within
the framework of a structured universal jet model (Zhang \&
Meszaros 2002a; Rossi et al. 2002). In this model, an observer
closer to the jet axis would detect a higher luminosity, thus at
higher redshift, smaller viewing angle detections are preferred
due to luminosity selection effect.

\begin{figure}
\centerline{\psfig{file=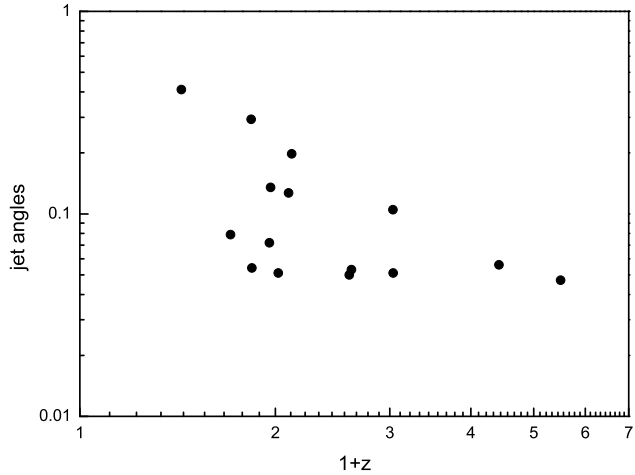,width=0.6\textwidth}}
\caption{The relation between the jet opening angle and redshift.
The data are taken from the paper of Frail et al. (2001). }
\end{figure}

It should be noted that in our analysis we have not considered the
errors coming from the luminosity - variability relation. We know
that the relation between luminosity and variability is somewhat
scatter, the correlation coefficient $n$ ($L\propto V^n$) can be
range from 2.2 to 5.8 (the best value is 3.3, see Fenimore \&
Ramirez-Ruiz 2001), so it is natural that the redshifts and
luminosities inferred from this luminosity - variability relation
should have large errors, and these errors should be somehow
transferred to the final errors in the correlation indices.
However this effect is very complicated, we hope that this effect
can be taken into account in the future work.

In summary, in this paper we discuss the properties of two group
GRBs, one group with known redshift from afterglow observation,
and another group with redshift derived from the luminosity -
variability relation. We find that the properties of these two
group GRBs are all correlated with the redshifts, which reinforces
the validity of the redshift estimates derived from the luminosity
- variability relation. If this is true, then we see that the
burst features, such as their intrinsic duration, luminosity,
radiated energy and peak energy $E_p$, are all redshift dependent,
which means that there are cosmological evolution effects on
gamma-ray bursts features, and this can provide an interesting
clue to the nature of GRBs.

\section{acknowledgements}
We are very grateful to the referee for several important comments
that greatly improved this paper. This work is supported by the
National Natural Science Foundation (grants 10073022, 10233010 and
10225314) and the National 973 Project on Fundamental Researches
of China (NKBRSF G19990754).

\end{document}